\documentclass[12pt,prd]{revtex4}
\usepackage{amssymb}
\usepackage{amsmath}

\input{tcilatex}

\begin{document}

\title{Spinning Up Asymptotically Flat Spacetimes }
\author{E.N. Glass* and J.P. Krisch}
\affiliation{Department of Physics, University of Michigan, Ann Arbor, Michgan 48109}
\thanks{Permanent address: Physics Department, University of Windsor,
Ontario N9B 3P4, Canada}
\date{22 September 2004 }

\begin{abstract}
\ \newline
We present a method for constructing stationary, asymptotically flat,
rotating solutions of Einstein's field equations. One of the spun-up
solutions has quasilocal mass but no global mass. It has an ergosphere but
no event horizon. The angular momentum is constant everywhere beyond the
ergosphere. The energy-momentum content of this solution can be interpreted
as a rotating string-fluid.\newline
\ \newline
PACS numbers: 04.20.Jb, 04.40.Dg
\end{abstract}

\maketitle

\section{INTRODUCTION}

A metric description of the spacetime around an object encodes information
about its mass and angular momentum. While many of the objects of interest
to astrophysicists and cosmologists are rotating, there are not a large
number of metrics that can be used to describe isolated rotating objects.
Most models of the early universe encode no angular momentum into the
metric. Developing methods to generate rotating solutions is clearly of
value. There are two well known solution generating algorithms which start
with a static metric and transform it by adding rotation to some aspect of
the spacetime structure. The complex transformation discovered by Newman
(N-J) and others \cite{NJ65},\cite{NCC+65},\cite{New73}, for example, takes
the Schwarzschild metric to the Kerr metric, and the Reissner-Nordstr\"{o}m
metric to the Kerr-Newman solution. The N-J transform adds angular momentum
to a seed spacetime initially containing only global mass (and possibly
charge). The N-J method has also been used \cite{DT97} to spin up spherical
metrics to obtain Kerr interiors. There is a method begun by Ehlers \cite%
{Ehl62} and fully developed by Geroch \cite{Ger71},\cite{Ger72} which adds
twist to the timelike Killing vector of the original (seed) metric and also
changes its norm. If the method is applied to the static Schwarzschild
spacetime the Taub-Nut spacetime results.\ 

In this paper, we suggest a new method for adding rotation to a known static
spacetime. The method starts with an asymptotically flat static metric
written in null Bondi-Sachs form. A transform of the type $d\tilde{\varphi}%
=d\varphi -\Omega du$ is applied.\ $\Omega $ has coordinate dependence thus
adding angular momentum to the spacetime. The Komar integral is used to
calculate the angular momentum. Asymptotic flatness is enforced by using the
Bondi-Sachs metric as a prototype for adding angular momentum, with the
Bondi-Sachs metric conditions ensuring that asymptotic flatness is preserved
in the transformation.

In the next section we describe the method in detail. In the third part of
the paper we apply the method to the\ Schwarzschild spacetime, adding a
rotation term, maintaining the Bondi-Sachs metric form, and using the
rotation term from the linearized Kerr metric in the Bondi-Sachs frame as a
guide. We also apply the method to Minkowski space, obtainable as a limiting
case of the spinning Schwarzschild example.\ This spacetime contains some
unusual features. Unusual because there is quasilocal mass but no global
mass, and there is an ergosphere but no event horizon. The solution is
stationary, axially symmetric, and asymptotically flat. The energy-momentum
content can be identified as a rotating string fluid. Values from the Komar
superpotential provide angular momentum which does not fall off
asymptotically and quasilocal mass which does. The quasilocal mass arises
from the kinetic energy of the rotating string fluid.

\textbf{CONVENTIONS}

In this work Greek indices range over (0,1,2,3) = ($u,r,\vartheta ,\varphi $%
). Sign conventions are $2A_{\nu ;[\alpha \beta ]}=A_{\mu }R_{\ \nu \alpha
\beta }^{\mu },$ and $R_{\alpha \beta }=R_{\ \alpha \beta \nu }^{\nu }.$ The
metric signature is (+,-,-,-) and the field equations are $G_{\mu \nu
}=-8\pi T_{\mu \nu }$.

\section{THE METHOD}

The Bondi-Sachs metric has been used since the 1960s. It is based on a
foliation of outgoing null hypersurfaces ($\mathcal{N}$) and a conformal
boundary ($\mathcal{I}$), with sufficient generality to describe bounded
radiating astrophysical systems (details are given in Appendix D).
Asymptotically flat electrovac spacetimes in Bondi-Sachs form are summarized
by Bicak and Pravdova \cite{BP98}.

To construct our solution, we begin with the Schwarzschild vacuum metric.
Angular momentum is added using the $\partial _{\varphi }$ Killing vector as
generator of rotations. This is done in the context of the Bondi-Sachs
metric in order to identify the term to add, and to maintain asymptotic
flatness. A new term is added containing angular velocity with form $d\tilde{%
\varphi}=d\varphi -\Omega du$. For $\Omega $ a function rather than a
constant, this is not an integrable coordinate transformation. The Komar
superpotential integral for $\partial _{\varphi }$ insures that the desired
amount of angular momentum is added. The particular choice of $\Omega $ is
guided by the linearized Kerr metric written in Bondi-Sachs form. We then
have an analytic asymptotically flat spinning metric.

\subsection{Kerr Vacuum Solution}

The Kerr solution in Boyer-Lindquist coordinates is given by 
\begin{eqnarray*}
g_{\alpha \beta }^{\text{kerr}}dx^{\alpha }dx^{\beta } &=&\psi \
dt^{2}-(\Sigma /\Delta )\ dr^{2}+(1-\psi )2a\text{sin}^{2}\vartheta \
dtd\varphi \\
&&-\Sigma \ d\vartheta ^{2}-\text{sin}^{2}\vartheta \lbrack \Sigma +(2-\psi
)a^{2}\text{sin}^{2}\vartheta ]\ d\varphi ^{2}
\end{eqnarray*}%
Here $\psi =1-2m_{0}r/\Sigma $, $\Sigma =r^{2}+a^{2}\cos ^{2}\vartheta $, $%
\Delta =r^{2}+a^{2}-2m_{0}r$. The linearized Kerr metric has terms linear in
'$a$'. 
\begin{eqnarray}
g_{\alpha \beta }^{\text{lin-kerr}}dx^{\alpha }dx^{\beta }
&=&(1-2m_{0}/r)dt^{2}-(1-2m_{0}/r)^{-1}dr^{2}+(2m_{0}a/r)\text{sin}%
^{2}\vartheta \ 2dtd\varphi  \label{lin-kerr} \\
&&-r^{2}(d\vartheta ^{2}+\sin ^{2}\vartheta )d\varphi ^{2}.  \notag
\end{eqnarray}%
To reach Bondi-Sachs form transform from Boyer-Lindquist '$t$' to outgoing $%
u=t-r-2m_{0}\ln (r-2m_{0})$. 
\begin{eqnarray}
g_{\alpha \beta }^{\text{lin-kerr-BS}}dx^{\alpha }dx^{\beta }
&=&(1-2m_{0}/r)du^{2}+2dudr+(2m_{0}a/r)\text{sin}^{2}\vartheta \ 2dud\varphi
\label{lin-kerr-bs} \\
&&-(1-2m_{0}/r)^{-1}(2m_{0}a/r)\text{sin}^{2}\vartheta \ 2drd\varphi
-r^{2}(d\vartheta ^{2}+\sin ^{2}\vartheta )d\varphi ^{2}.  \notag
\end{eqnarray}%
Note that 
\begin{equation*}
(du)_{\alpha }(du)_{\beta }\ g_{\text{kerr}}^{\alpha \beta }=0\text{ to
order '}a\text{'.}
\end{equation*}

\section{SPINNING SCHWARZSCHILD}

The Schwarzschild vacuum is spun up by choosing a term with angular velocity 
$\Omega (r)$. The spinning metric is written in Bondi-Sachs form as 
\begin{equation}
g_{\alpha \beta }^{\text{BS}}dx^{\alpha }dx^{\beta
}=(1-2m_{0}/r)du^{2}+2dudr-r^{2}d\vartheta ^{2}-r^{2}\sin ^{2}\vartheta
\,(d\varphi -\Omega du)^{2}.  \label{bsmet}
\end{equation}%
The term $(2m_{0}a/r)$sin$^{2}\vartheta \ 2dud\varphi $ in the linearized
Kerr metric Eq.(\ref{lin-kerr-bs}) leads to the function $U^{\varphi }=$ $%
\Omega (r)$ in Eq.(\ref{bsmetric}), multiplied by $r^{2}H_{\varphi \varphi
}=r^{2}\sin ^{2}\vartheta $. This directs the choice $\Omega
(r)=r_{0}^{2}/r^{3}$.

The Bondi-Sachs form can be rewritten as the spinning Schwarzschild (sSCH)
metric%
\begin{equation}
g_{\alpha \beta }^{\text{sSCH}}dx^{\alpha }dx^{\beta }=Zdu^{2}+2dudr+2\Omega
r^{2}\text{sin}^{2}\vartheta \,dud\varphi -r^{2}(d\vartheta ^{2}+\sin
^{2}\vartheta \,d\varphi ^{2})  \label{ssch-met}
\end{equation}%
with\ $Z:=1-2m_{0}/r-(\Omega r$sin$\vartheta )^{2}$.

To understand the physical content of the sSCH solution we write the metric
in a basis which is locally non-rotating ($\hat{v}^{\alpha }$ has zero
twist). 
\begin{equation}
g_{\alpha \beta }^{\text{sSCH}}=\hat{v}_{\alpha }\hat{v}_{\beta }-\hat{r}%
_{\alpha }\hat{r}_{\beta }-\hat{\vartheta}_{\alpha }\hat{\vartheta}_{\beta }-%
\hat{\varphi}_{\alpha }\hat{\varphi}_{\beta }.  \label{ssch_tet}
\end{equation}%
Here $A^{2}=1-2m_{0}/r$. The unit vectors are defined by 
\begin{subequations}
\label{ssch-tet}
\begin{eqnarray}
\hat{v}_{\alpha }dx^{\alpha } &=&A^{-1}(A^{2}du+dr),\ \ \ \ \ \ \ \hat{v}%
^{\alpha }\partial _{\alpha }=A^{-1}(\partial _{u}+\Omega \partial _{\varphi
}),  \label{tetv} \\
\hat{r}_{\alpha }dx^{\alpha } &=&A^{-1}dr,\ \ \ \ \ \ \ \ \ \ \ \ \ \ \ \ \
\ \ \ \hat{r}^{\alpha }\partial _{\alpha }=A^{-1}(\partial
_{u}-A^{2}\partial _{r}+\Omega \partial _{\varphi }),  \label{tetr} \\
\hat{\vartheta}_{\alpha }dx^{\alpha } &=&rd\vartheta ,\ \ \ \ \ \ \ \ \ \ \
\ \ \ \ \ \ \ \ \ \ \ \ \hat{\vartheta}^{\alpha }\partial _{\alpha
}=-r^{-1}\partial _{\vartheta },  \label{tetth} \\
\hat{\varphi}_{\alpha }dx^{\alpha } &=&r\text{sin}\vartheta (d\varphi
-\Omega du),\ \ \ \ \ \ \hat{\varphi}^{\alpha }\partial _{\alpha }=-(r\text{%
sin}\vartheta )^{-1}\partial _{\varphi }.  \label{tetph}
\end{eqnarray}%
The Einstein tensor is expanded as 
\end{subequations}
\begin{equation}
G_{\alpha \beta }^{\text{sSCH}}=U^{2}(\hat{v}_{\alpha }\hat{v}_{\beta }-\hat{%
r}_{\alpha }\hat{r}_{\beta }+\hat{\vartheta}_{\alpha }\hat{\vartheta}_{\beta
}+3\hat{\varphi}_{\alpha }\hat{\varphi}_{\beta })  \label{ssch-ein}
\end{equation}%
with $U^{2}=(9/4)\Omega ^{2}\sin ^{2}\vartheta $. In the locally
non-rotating frame, the energy-momentum tensor is 
\begin{equation}
T_{\alpha \beta }^{\text{sSCH}}=\rho \hat{v}_{\alpha }\hat{v}_{\beta }+p_{r}%
\hat{r}_{\alpha }\hat{r}_{\beta }+p_{\vartheta }\hat{\vartheta}_{\alpha }%
\hat{\vartheta}_{\beta }+p_{\varphi }\hat{\varphi}_{\alpha }\hat{\varphi}%
_{\beta }.
\end{equation}%
It follows from $G_{\alpha \beta }=-8\pi T_{\alpha \beta }$ that 
\begin{eqnarray}
8\pi \rho &=&-U^{2},\ \ \ 8\pi p_{r}=U^{2}, \\
8\pi p_{\vartheta } &=&-U^{2},\ \ \ 8\pi p_{\varphi }=-3U^{2}.  \notag
\end{eqnarray}%
The Ricci tensor has the form 
\begin{equation}
R_{\alpha \beta }^{\text{sSCH}}=2U^{2}(\hat{v}_{\alpha }\hat{v}_{\beta }-%
\hat{r}_{\alpha }\hat{r}_{\beta }+\hat{\varphi}_{\alpha }\hat{\varphi}%
_{\beta }),\ \ R_{\alpha \beta }^{\text{sSCH}}l^{\beta }=2U^{2}l_{\alpha },
\end{equation}%
with $l^{\alpha }\partial _{\alpha }=\partial _{r}$ an eigenvector of the
Ricci tensor.

The fluid velocity $\hat{v}^{\alpha }$ is expansion-free and twist-free,
with shear scalar $\frac{1}{2}\sigma _{\alpha \beta }\sigma ^{\alpha \beta
}=U^{2}$, and acceleration $a^{\alpha }=(\frac{m_{0}}{Ar^{2}})\hat{r}%
^{\alpha }$. The content of spinning Schwarzschild can be interpreted as a
rotating radial string fluid with $\rho +p_{r}=0$.

The sSCH metric can be written as 
\begin{equation}
g_{\alpha \beta }^{\text{sSCH}}=g_{\alpha \beta }^{\text{SCH}}-2\Omega
l_{\alpha }S_{\beta }\text{,}  \label{ssch-sch}
\end{equation}%
where $g_{\alpha \beta }^{\text{SCH}}$ is the Schwarzschild metric and $%
S_{\beta }$ is a spacelike vector such that 
\begin{eqnarray}
S_{\beta }dx^{\beta } &=&(\Omega /2)r^{2}\sin ^{2}\vartheta du-r^{2}\sin
^{2}\vartheta d\varphi , \\
S_{\alpha }S^{\alpha } &=&-(r\sin \vartheta )^{2}.  \notag
\end{eqnarray}%
When $m_{0}\rightarrow 0$ in sSCH, we have spinning Minkowski (sM): 
\begin{equation}
g_{\alpha \beta }^{\text{sM}}=\eta _{\alpha \beta }-2\Omega l_{\alpha
}S_{\beta }.  \label{smk-ks}
\end{equation}%
This spacetime, as seen from null infinity, has angular momentum but no mass.

\subsection{Spinning Minkowski}

We set $m_{0}=0$ in $Z$ of Eq.(\ref{ssch-met}) to obtain a metric with no
global mass. The spinning Minkowski metric (sM) has one-parameter, $r_{0}$,
in function $\Omega (r)=r_{0}^{2}/r^{3}$. 
\begin{equation}
g_{\alpha \beta }^{\text{sM}}dx^{\alpha }dx^{\beta }=(1-\Omega ^{2}r^{2}%
\text{sin}^{2}\vartheta )du^{2}+2dudr+2\Omega r^{2}\text{sin}^{2}\vartheta
\,dud\varphi -r^{2}(d\vartheta ^{2}+\sin ^{2}\vartheta \,d\varphi ^{2}).
\label{sm-met}
\end{equation}%
The metric has two Killing vectors, stationary $k_{\text{(u)}}$ and axial $%
k_{\text{(}\varphi \text{)}}.$ In terms of the Minkowski metric $\eta
_{\alpha \beta }$, the sM metric can be written as%
\begin{equation}
g_{\alpha \beta }^{\text{sM}}dx^{\alpha }dx^{\beta }=\eta _{\alpha \beta
}dx^{\alpha }dx^{\beta }-\Omega ^{2}r^{2}\text{sin}^{2}\vartheta
du^{2}+2\Omega r^{2}\text{sin}^{2}\vartheta \,dud\varphi .  \notag
\end{equation}%
The rotation axis, [$\vartheta $: $0,2\pi $] is a regular line in the
spacetime. As $r\rightarrow \infty $, $\Omega r^{2}\rightarrow 0$ and $%
g_{\alpha \beta }^{\text{sM}}\rightarrow \eta _{\alpha \beta }$, hence the
sM metric is asymptotically flat. \ In fact, one can remap the Minkowski
metric 
\begin{equation*}
\eta _{\alpha \beta }dx^{\alpha }dx^{\beta }=du^{2}+2dudr-r^{2}(d\vartheta
^{2}+\sin ^{2}\vartheta \,d\tilde{\varphi}^{2})
\end{equation*}%
by shifting the $\tilde{\varphi}$ coordinate to $\tilde{\varphi}=\varphi
-\Omega _{0}u$. 
\begin{equation*}
\eta _{\alpha \beta }dx^{\alpha }dx^{\beta }=(1-\Omega _{0}^{2}r^{2}\text{sin%
}^{2}\vartheta )du^{2}+2dudr+2\Omega _{0}r^{2}\text{sin}^{2}\vartheta
\,dud\varphi -r^{2}(d\vartheta ^{2}+\sin ^{2}\vartheta \,d\varphi ^{2}).
\end{equation*}%
The Minkowski metric then coincides with the limit of the sM metric as $%
{}\Omega (_{r\rightarrow \infty }^{\text{lim}})\rightarrow 0$, and $\Omega
_{0}\rightarrow 0$.

The sM solution is written in the same locally non-rotating frame as sSCH
above, but now the frame velocity $\hat{v}^{\alpha }$ is geodesic. It is
also expansion-free and twist-free, with shear scalar $\frac{1}{2}\sigma
_{\alpha \beta }\sigma ^{\alpha \beta }=U^{2}=(3r_{0}^{2}\sin \vartheta
/2r^{3})^{2}$. 
\begin{equation}
g_{\alpha \beta }^{\text{sM}}=\hat{v}_{\alpha }\hat{v}_{\beta }-\hat{r}%
_{\alpha }\hat{r}_{\beta }-\hat{\vartheta}_{\alpha }\hat{\vartheta}_{\beta }-%
\hat{\varphi}_{\alpha }\hat{\varphi}_{\beta }.
\end{equation}%
The unit vectors are [this is the tetrad in Eq.(\ref{ssch-tet}) with $A=1$] 
\begin{subequations}
\label{v-tetrad}
\begin{eqnarray}
\hat{v}_{\alpha }dx^{\alpha } &=&du+dr,\ \ \ \ \ \ \ \ \ \ \ \ \ \ \ \ \ \ 
\hat{v}^{\alpha }\partial _{\alpha }=\partial _{u}+\Omega \partial _{\varphi
},  \label{v-hat} \\
\hat{r}_{\alpha }dx^{\alpha } &=&dr,\ \ \ \ \ \ \ \ \ \ \ \ \ \ \ \ \ \ \ \
\ \ \ \ \ \hat{r}^{\alpha }\partial _{\alpha }=\partial _{u}-\partial
_{r}+\Omega \partial _{\varphi },  \label{r-hat} \\
\hat{\vartheta}_{\alpha }dx^{\alpha } &=&rd\vartheta ,\ \ \ \ \ \ \ \ \ \ \
\ \ \ \ \ \ \ \ \ \ \ \ \hat{\vartheta}^{\alpha }\partial _{\alpha
}=-r^{-1}\partial _{\vartheta },  \label{theta-hat} \\
\hat{\varphi}_{\alpha }dx^{\alpha } &=&r\text{sin}\vartheta (d\varphi
-\Omega du),\ \ \ \ \ \ \hat{\varphi}^{\alpha }\partial _{\alpha }=-(r\text{%
sin}\vartheta )^{-1}\partial _{\varphi }.  \label{phi-hat}
\end{eqnarray}

The Einstein tensor and energy-momentum tensor here are the same as those of
the sSCH solution above. 
\end{subequations}
\begin{equation}
G_{\alpha \beta }^{\text{sM}}=U^{2}(\hat{v}_{\alpha }\hat{v}_{\beta }-\hat{r}%
_{\alpha }\hat{r}_{\beta }+\hat{\vartheta}_{\alpha }\hat{\vartheta}_{\beta
}+3\hat{\varphi}_{\alpha }\hat{\varphi}_{\beta }).  \label{ein-sm}
\end{equation}%
\begin{equation}
T_{\alpha \beta }^{\text{sM}}=\rho \hat{v}_{\alpha }\hat{v}_{\beta }+p_{r}%
\hat{r}_{\alpha }\hat{r}_{\beta }+p_{\vartheta }\hat{\vartheta}_{\alpha }%
\hat{\vartheta}_{\beta }+p_{\varphi }\hat{\varphi}_{\alpha }\hat{\varphi}%
_{\beta },
\end{equation}%
\begin{eqnarray}
8\pi \rho &=&-U^{2},\ \ \ 8\pi p_{r}=U^{2}, \\
8\pi p_{\vartheta } &=&-U^{2},\ \ \ 8\pi p_{\varphi }=-3U^{2}.  \notag
\end{eqnarray}%
The sM solution contains a rotating radial string fluid with $\rho +p_{r}=0$%
. Since $U=(3/2)r_{0}^{2}\sin \vartheta /r^{3}$, the entire matter content
approaches vacuum as $r\rightarrow \infty $.

\section{Mass}

The Bondi mass of $g^{\text{sSCH}}$ is evaluated on a spherical cut of
future null infinity $\mathcal{I}^{+}$. Equation (\ref{bondimass}) and the
Weyl tensor component Eq.(\ref{ssch-psi2}) yields%
\begin{equation}
M_{bondi}^{\text{sSCH}}=-\frac{1}{8\pi }\doint\limits_{\partial \mathcal{N}%
}(\Psi _{2}^{{\small 0}}+\bar{\Psi}_{2}^{{\small 0}})\,\sqrt{-g}d\vartheta
d\varphi =m_{0}.  \notag
\end{equation}%
Since metric $g^{\text{sM}}$ is the $m_{0}\rightarrow 0$ limit of $g^{\text{%
sSCH}}$ it follows that 
\begin{equation}
M_{bondi}^{\text{sM}}=0.  \label{m-bondi}
\end{equation}

The Komar superpotential for Killing vector $k^{\beta }$ is 
\begin{equation}
U^{\alpha \beta }(k)=(-g)^{\frac{1}{2}}[\nabla ^{\alpha }k^{\beta }-\nabla
^{\beta }k^{\alpha }].
\end{equation}%
For timelike Killing vector $k_{\text{(u)}}^{\beta }\partial _{\beta
}=\partial _{u}$ one writes $U^{\alpha \beta }(\partial _{u})$, and for
axial Killing vector $k_{\text{(}\varphi \text{)}}^{\beta }\partial _{\beta
}=\partial _{\varphi }$ one writes $U^{\alpha \beta }(\partial _{\varphi })$%
. Metrics $g^{\text{sSCH}}$ and $g^{\text{sM}}$ both have $(-g)^{\frac{1}{2}%
}=r^{2}\sin \vartheta $. For $u=const$ three-surface $\mathcal{N}$, the mass
within a $u=const$, $r=const$ two-surface $\partial \mathcal{N}$ is 
\begin{equation}
M_{komar}=-\frac{1}{8\pi }\oint\limits_{\partial \mathcal{N}}U^{\alpha \beta
}(\partial _{u})\,dS_{\alpha \beta }.
\end{equation}%
Metric $g_{\alpha \beta }^{\text{sSCH}}$ and two-surface $dS_{\alpha \beta
}=u,_{[\alpha }r,_{\beta ]}d\vartheta d\varphi $ contains global mass $m_{0}$
and quasilocal mass at all $r$ beyond the ergosphere. 
\begin{eqnarray}
M_{komar}^{\text{sSCH}} &=&-\frac{1}{8\pi }\dint\limits_{0}^{2\pi
}\dint\limits_{0}^{\pi }(-2m_{0}\sin \vartheta -3\frac{r_{0}^{4}}{r^{3}}\sin
^{3}\vartheta )d\vartheta d\varphi =m_{0}+\frac{r_{0}^{4}}{r^{3}},
\label{m-komar} \\
M_{komar}^{\text{sM}} &=&\frac{r_{0}^{4}}{r^{3}}.  \notag
\end{eqnarray}%
There is no global mass for $g^{\text{sM}}$ since $m_{0}=0$ and $M_{komar}^{%
\text{sM}}\rightarrow 0$ as $r\rightarrow \infty $.

Although the sM solution is not spherically symmetric, we can compute a
sectional curvature mass since tetrad vectors $\hat{\vartheta}^{\alpha }$
and $\hat{\varphi}^{\alpha }$ of Eq.(\ref{theta-hat}) and (\ref{phi-hat})
form a bivector which satisfies the Frobenius suface-forming condition. They
span a family of closed two-surfaces. The sectional curvature (Gaussian
curvature) mass is 
\begin{equation}
-\frac{2M_{curv}^{\text{sSCH}}}{r^{3}}=R_{\alpha \beta \mu \nu }\hat{%
\vartheta}^{\alpha }\hat{\varphi}^{\beta }\hat{\vartheta}^{\mu }\hat{\varphi}%
^{\nu }=m_{0}.  \label{m-curv}
\end{equation}%
The sectional curvature mass for $g^{\text{sM}}$ vanishes.

The explanation of why $M^{\text{sSCH}}$ and $M^{\text{sM}}$ are positive
when the density $\rho =-U^{2}/8\pi $ is negative lies with the relativistic
contribution of pressure to mass. If one traces the Komar integral back it
then becomes clear. 
\begin{eqnarray*}
M_{komar} &=&-\frac{1}{8\pi }\oint\limits_{\partial \mathcal{N}}U^{\alpha
\beta }(\partial _{u})\,dS_{\alpha \beta } \\
&=&-\frac{1}{8\pi }\oint\limits_{\partial \mathcal{N}}\sqrt{-g}k_{\text{(u)}%
}^{[\alpha ;\beta ]}\,dS_{\alpha \beta } \\
&=&-\frac{1}{8\pi }\dint\limits_{\mathcal{N}}\sqrt{-g}k_{\text{(u)}}^{\alpha
}R_{\alpha }^{\ \beta }dS_{\beta }
\end{eqnarray*}%
for $dS_{\beta }=u_{,\beta }d^{3}x$. The three-volume integration is over
any $u=const$ null surface from $r$ (beyond the ergosurface) to $\infty $.
The Ricci tensor provides 
\begin{eqnarray*}
M_{komar} &=&-\frac{1}{8\pi }\dint\limits_{\mathcal{N}}\sqrt{-g}(-2\rho
-p_{r}+p_{\varphi })d^{3}x \\
&=&\frac{1}{4\pi }\dint\limits_{\mathcal{N}}\sqrt{-g}U^{2}drd\vartheta
d\varphi .
\end{eqnarray*}%
For $g^{\text{sSCH}}$ the volume integral does not measure the $m_{0}$
contribution.

\section{Angular momentum}

For $u=const$ three-surface $\mathcal{N}$, the angular momentum within a $%
u=const$, $r=const$ two-surface $\partial \mathcal{N}$ is 
\begin{equation}
J=-\frac{1}{16\pi }\oint\limits_{\partial \mathcal{N}}U^{\alpha \beta
}(\partial _{\varphi })\,dS_{\alpha \beta }.
\end{equation}%
Metric $g^{\text{sSCH}}$ and metric $g^{\text{sM}}$ admit constant $J$ at
all $r$ beyond the ergosphere. For two-surface element $dS_{\alpha \beta
}=u,_{[\alpha }r,_{\beta ]}d\vartheta d\varphi $%
\begin{equation}
J^{\text{sSCH}}=J^{\text{sM}}=-\frac{1}{16\pi }\dint\limits_{0}^{2\pi
}\dint\limits_{0}^{\pi }(3r_{0}^{2}\sin ^{3}\vartheta )d\vartheta d\varphi =-%
\frac{1}{2}r_{0}^{2}.  \label{kom-angmom}
\end{equation}%
This is a global value.

The normalization of the Komar integrals is chosen \cite{GN94} so that 
\begin{equation*}
J^{\text{kerr}}(\partial _{\varphi })=-m_{0}a\ \text{\ \ and }\ \ M^{\text{%
kerr}}(\partial _{u})=m_{0}.
\end{equation*}%
The sign difference occurs because our $-2$ signature gives timelike vectors
positive norms and spacelike vectors negative norms.

\section{SURFACES OF sM}

\subsection{Horizon}

Spacelike closed two-surfaces are trapped if their null generators (incoming
and outgoing null geodesics) converge toward the future. The time evolution
of the outermost trapped surface is a null surface called the "apparent
horizon". The spherical two-surfaces of metric $g^{\text{sM}}$ have area $%
4\pi r^{2}$, with null geodesic generators $l^{\alpha }$ and $n^{\alpha }$
[see Eq.(\ref{sm-np})]. Their expansions (convergences) are given
respectively by 
\begin{eqnarray}
\rho &=&-l^{\alpha }\partial _{\alpha }(\ln r)=-1/r, \\
\mu &=&n^{\alpha }\partial _{\alpha }(\ln r)=-1/(2r).  \notag
\end{eqnarray}%
(here $\rho $ is a spin-coefficient, not mass-density). There can be no
horizon since neither $\rho $ nor $\mu $ change sign over the interval $%
0<r<\infty $.

\subsection{Ergosurface}

The norm of the timelike Killing vector is, with $\Omega =r_{0}^{2}/r^{3}$, 
\begin{equation}
\partial _{u}^{\text{sM}}\cdot \partial _{u}^{\text{sM}}=1-\Omega
^{2}r^{2}\sin ^{2}\vartheta ,
\end{equation}%
placing the ergosphere at $r^{2}=\pm r_{0}^{2}\sin \vartheta =\mid
r_{0}^{2}\sin \vartheta \mid $.

The shape of the ergosurface (outer boundary of the ergosphere) is
determined by the locus of points where the norm of $\partial _{u}^{\text{sM}%
}$ is zero, $r^{2}=\mid r_{0}^{2}\sin \vartheta \mid $. The curve ranges
from $r=0$ to $r=r_{0}$, and is called the \textquotedblright Leminscate of
Bernoulli\textquotedblright\ \cite{Law72}. In ($x,y,z$) coordinates, with
rotation about the z-axis, the ergosurface is shaped like a deformed torus
without a center hole, since the surface boundary curve goes smoothly to $0$
at $x=y=z=0$. Inside the ergosurface the spacelike two-surfaces are
two-spheres.

With $\partial _{u}^{\text{sM}}\cdot \partial _{u}^{\text{sM}}<0$, the
Gaussian curvature of $\vartheta $,$\varphi $ two-surfaces inside the
ergosphere is 
\begin{equation}
\mathcal{K}_{\text{ergo}}=1/r^{2}.  \label{ergo-curv}
\end{equation}%
We compute the Euler-Poincar\'{e} characteristic $\chi $ of the ergosurface.
The two-surface metric is 
\begin{equation}
ds_{\text{2-surf}}^{2}=r_{0}^{2}\sin \vartheta (d\vartheta ^{2}+\sin
^{2}\vartheta \,d\varphi ^{2})
\end{equation}%
and 
\begin{equation}
\mathcal{K}_{\text{2-ergo}}=\frac{1}{r_{0}^{2}\sin \vartheta }
\end{equation}%
undefined at the poles $\vartheta =0,\pi $. The Euler-Poincar\'{e}
characteristic \cite{Car76} is then 
\begin{equation}
2\pi \chi =\int\limits_{0}^{2\pi }\int\limits_{0}^{\pi }\mathcal{K}_{\text{%
2-ergo}}(r_{0}^{2}\sin ^{2}\vartheta )\,d\vartheta d\varphi =4\pi .
\end{equation}%
Thus $\chi =2$, with associated genus $g=0$ which characterizes the topology
of a standard two-sphere.

The Kerr ergosphere is quite different. The norm of the Kerr timelike
Killing vector (in Boyer-Lindquist coordinates) is 
\begin{equation*}
\partial _{t}^{\text{Kerr}}\cdot \partial _{t}^{\text{Kerr}}=1-\frac{2m_{0}r%
}{r^{2}+a^{2}\cos ^{2}\vartheta }.
\end{equation*}%
The Killing vector is spacelike in the region $2m_{0}\leq
r<m_{0}+(m_{0}^{2}-a^{2}\cos ^{2}\vartheta )^{1/2}$. The Kerr ergosurface is
tangent to the rotating trapped surface at the poles.

\subsection{Geodesic Deviation of sM}

Test particles move along a congruence of timelike geodesics $v^{\alpha
}=dx^{\alpha }(\tau )/d\tau $. Vector $v^{\alpha }$ is twist-free and
acceleration-free with respect to metric $g^{\text{sM}}$ and is given by $%
v^{\alpha }\partial _{\alpha }=\partial _{u}+\Omega \partial _{\varphi }$.
Covariant differentiation along the geodesics is defined by 
\begin{equation}
D/d\tau :=v^{\alpha }\nabla _{\alpha }.
\end{equation}%
Deviation vector $\eta ^{\alpha }$ is tangent to the line connecting pairs
of neighboring geodesics in the congruence. It is Lie transported along the
congruence and satisfies 
\begin{equation}
\mathcal{L}_{v}\eta ^{\alpha }=0.
\end{equation}%
$\frac{D^{2}}{d\tau ^{2}}\eta ^{\alpha }$ gives the relative acceleration
between neighboring test particles. The geodesic deviation equation is \cite%
{Syn64},\cite{Sze65} 
\begin{equation}
\frac{D^{2}}{d\tau ^{2}}\eta ^{\alpha }=[R^{\alpha }{}_{\beta \mu \nu
}\,v^{\beta }v^{\mu }]\eta ^{\nu }.  \label{gde-eqn}
\end{equation}%
If one considers any $\varphi =const$ slice of a family of \{$\vartheta
,\varphi $\} two-spheres then $\eta =(1/\sin \vartheta )\partial _{\varphi
}\ $is a deviation vector linking timelike geodesics orthogonal to the
two-spheres. We find 
\begin{equation}
\frac{D^{2}}{d\tau ^{2}}\eta ^{\alpha }=(\frac{9}{4}\Omega ^{2}r)\hat{\varphi%
}^{\alpha }
\end{equation}%
for $\Omega =r_{0}^{2}/r^{3}$ and tetrad vector (\ref{phi-hat}). A view of
the geodesic congruence shows the tip of the connecting vector $\eta $
spiraling up the congruence with acceleration $\Omega ^{2}r$.

\section{DISCUSSION}

A method has been presented for adding angular momentum to static
spacetimes. Metric $g^{\text{sSCH}}$ is a spinning generalization of the
vacuum Schwarzschild metric. The spinup method creates a spinning object
with an atmosphere in an asymptotically flat spacetime. There are many
isolated astrophysical systems in our observable universe that spin. We have
recently seen that the unexpected behavior of the angular velocity of
globular clusters was a strong clue to the existence of dark matter and dark
energy. Analytic metrics, with a family of possible angular velocities, can
be a useful tool for studying and classifying rotating systems. They can be
used to predict and analyze observable data such as lensing and to check
numerical simulations.

Here we started with the Schwarzschild metric and added a rotation term,
maintaining Bondi-Sachs metric form and using the rotation term from the
linearized Kerr metric (transformed to a Bondi-Sachs frame) as a guide. The
resulting metric, $g^{\text{sSCH}}$, has global mass and angular momentum.
The global mass is $M_{bondi}^{\text{sSCH}}=m_{0}$, the Bondi mass. The
Komar mass of spinning Schwarzschild $M_{komar}^{\text{sSCH}%
}=m_{0}+r_{0}^{4}/r^{3}$ has both global mass and a quasilocal term which
falls off asymptotically. When $m_{0}$ is set to zero, then $g^{\text{sSCH}%
}\rightarrow g^{\text{sM}}$ the spinning Minkowski metric.

The $g^{\text{sM}}$ metric has unusual properties. Unusual because there is
quasilocal mass but no global mass, and there is an ergosphere but no event
horizon. Values from the Komar superpotential provide global angular
momentum, $J^{\text{sM}}=-\frac{1}{2}r_{0}^{2}$. The quasilocal mass does
fall off asymptotically. It arises from the kinetic energy of the rotating
string fluid. As seen from conformal infinity, $g^{\text{sM}}$ is a large
spinning object with no total weight.

In this first work we focused on asymptotically flat seed metrics, and spun
them up in a prescribed manner. Future work will allow other configurations,
such as cylindrical symmetry, or extended matter, and the Bondi-Sachs metric
form can be relaxed.

\appendix

\section{NULL TETRAD}

\subsection{Spinning Schwarzschild}

The null tetrad which spans metric $g_{\alpha \beta }^{\text{sSCH}}$ of Eq.(%
\ref{ssch-met}) is, with $\Omega =r_{0}^{2}/r^{3}$ and $A^{2}=1-2m_{0}/r$, 
\begin{eqnarray}
l_{\mu }dx^{\mu } &=&du,\ \ \ \ \ \ \ \ \ \ \ \ \ \ \ \ \ \ \ \ \ \ \ \ \ \
\ \ \ \ \ \ \ \ \ \ \ \ \ \ \ \ \ \ \ \ \ \ \ l^{\mu }\partial _{\mu
}=\partial _{r},  \label{ssch-np} \\
n_{\mu }dx^{\mu } &=&(A^{2}/2)du+dr,\ \ \ \ \ \ \ \ \ \ \ \ \ \ \ \ \ \ \ \
\ \ \ \ \ \ \ \ \ \ \ \ \ n^{\mu }\partial _{\mu }=\partial
_{u}-(A^{2}/2)\partial _{r}+\Omega \partial _{\varphi },  \notag \\
m_{\mu }dx^{\mu } &=&(\frac{i\,\Omega r\sin \vartheta }{\sqrt{2}})du-(\frac{r%
}{\sqrt{2}})(d\vartheta +i\text{sin}\vartheta \,d\varphi ),\ \ \ m^{\mu
}\partial _{\mu }=\frac{1}{\sqrt{2}r}(\partial _{\vartheta }+\frac{i}{\sin
\vartheta }\partial _{\varphi }).  \notag
\end{eqnarray}%
The Weyl tensor components are%
\begin{eqnarray}
\Psi _{0} &=&0,  \notag \\
\Psi _{1} &=&(\frac{3i}{2\sqrt{2}})\frac{\Omega \sin \vartheta }{r},  \notag
\\
\Psi _{2} &=&-\frac{m_{0}}{r^{3}}-\frac{3}{2}(\Omega ^{2}\sin ^{2}\vartheta
+i\,\frac{\Omega }{r}\cos \vartheta ),  \label{ssch-psi2} \\
\Psi _{3} &=&-\frac{\Psi _{1}}{2}(1-\frac{2m_{0}}{r}),  \notag \\
\Psi _{4} &=&0.  \notag
\end{eqnarray}%
The Ricci components are 
\begin{eqnarray*}
\Phi _{00} &=&\Phi _{01}=\Phi _{12}=\Phi _{22}=0, \\
\Phi _{02} &=&\frac{9}{8}\Omega ^{2}\sin ^{2}\vartheta , \\
\Phi _{11} &=&-\frac{3}{2}\Phi _{02}, \\
6\Lambda &=&\mathcal{R}/4=\Phi _{02}.
\end{eqnarray*}

\subsection{Spinning Minkowksi}

The massless spinning metric, with $\Omega =r_{0}^{2}/r^{3}$ 
\begin{equation*}
g_{\alpha \beta }^{\text{sM}}dx^{\alpha }dx^{\beta }=(1-\Omega ^{2}r^{2}%
\text{sin}^{2}\vartheta )du^{2}+2dudr+2\Omega r^{2}\text{sin}^{2}\vartheta
\,dud\varphi -r^{2}(d\vartheta ^{2}+\sin ^{2}\vartheta \,d\varphi ^{2})
\end{equation*}%
is expanded in a null tetrad as 
\begin{equation}
g_{\alpha \beta }^{\text{sM}}=l_{\alpha }n_{\beta }+n_{\alpha }l_{\beta
}-m_{\alpha }\bar{m}_{\beta }-\bar{m}_{\alpha }m_{\beta }  \label{sm-np-tet}
\end{equation}%
where%
\begin{eqnarray}
l_{\alpha }dx^{\alpha } &=&du,\ \ \ \ \ \ \ \ \ \ \ \ \ \ \ \ \ \ \ \ \ \ \
\ \ \ \ \ \ \ \ \ \ \ \ l^{\alpha }\partial _{\alpha }=\partial _{r},
\label{sm-np} \\
n_{\alpha }dx^{\alpha } &=&du/2+dr,\text{ \ \ \ \ \ \ \ \ }\ \ \ \text{\ \ \
\ \ \ \ \ \ \ \ \ \ }n^{\alpha }\partial _{\alpha }=\partial
_{u}-(1/2)\partial _{r}+\Omega \partial _{\varphi },  \notag \\
m_{\alpha }dx^{\alpha } &=&\frac{i}{\sqrt{2}}(\Omega r\sin \vartheta )du-%
\frac{r}{\sqrt{2}}(d\vartheta +i\text{sin}\vartheta \,d\varphi ),\text{ \ }%
m^{\alpha }\partial _{\alpha }=\frac{1}{\sqrt{2}r}(\partial _{\vartheta }+%
\frac{i}{\sin \vartheta }\partial _{\varphi })  \notag
\end{eqnarray}%
This tetrad has 6 zero-valued spin coefficients: $\kappa =\epsilon =\sigma
=\lambda =\nu =\gamma =0$. The others are 
\begin{eqnarray*}
\rho  &=&-\frac{1}{r},\text{ \ \ \ \ \ \ \ \ \ \ \ \ \ \ \ \ \ \ \ \ \ \ \ \
\ \ \ \ \ \ \ \ \ }\mu =-\frac{1}{2r}, \\
\pi  &=&(\frac{3i}{2\sqrt{2}})\Omega \sin \vartheta ,\text{ \ \ \ \ \ \ \ \
\ \ \ \ \ \ \ \ \ \ \ }\tau =-\pi , \\
\beta  &=&\frac{1}{2\sqrt{2}}\left( \frac{\cot \vartheta }{r}-\frac{3i}{2}%
\Omega \sin \vartheta \right) ,\text{ \ \ \ }\alpha =-\beta .
\end{eqnarray*}%
$l^{\alpha }$is geodesic since $\kappa =0,$ and $n^{\alpha }$ is geodesic
since $\nu =0$.

The Weyl tensor components are 
\begin{eqnarray}
\Psi _{0} &=&0,  \notag \\
\Psi _{1} &=&(\frac{3i}{2\sqrt{2}})\frac{\Omega \sin \vartheta }{r},  \notag
\\
\Psi _{2} &=&-\frac{3}{2}(\Omega ^{2}\sin ^{2}\vartheta +i\,\frac{\Omega }{r}%
\cos \vartheta ),  \label{sm-psi2} \\
\Psi _{3} &=&-\Psi _{1}/2,  \notag \\
\Psi _{4} &=&0.  \notag
\end{eqnarray}%
If there was a Bondi mass, it would appear in $\Psi _{2}$ at $O(1/r^{3})$
since $r$ is a valid luminosity distance. The Einstein tensor null tetrad
components are 
\begin{equation}
G_{\alpha \beta }^{\text{sM}}=-4\Phi _{11}(l_{\alpha }n_{\beta }+n_{\alpha
}l_{\beta })-2\Phi _{02}(m_{\alpha }m_{\beta }+\bar{m}_{\alpha }\bar{m}%
_{\beta })+(2\Phi _{11}-\mathcal{R}/4)g_{\alpha \beta }^{\text{sM}}
\end{equation}%
where 
\begin{eqnarray*}
\Phi _{00} &=&\Phi _{01}=\Phi _{12}=\Phi _{22}=0, \\
\Phi _{02} &=&\frac{9}{8}\Omega ^{2}\sin ^{2}\vartheta =\frac{U^{2}}{2}, \\
\Phi _{11} &=&-\frac{3}{2}\Phi _{02}, \\
6\Lambda &=&\mathcal{R}/4=\Phi _{02}.
\end{eqnarray*}%
One can transform from the null tetrad to the locally non-rotating tetrad in
Eq.(\ref{v-tetrad}): 
\begin{equation}
\hat{v}=l/2+n,\ \ \hat{r}=-l/2+n,\ \ \hat{\vartheta}=-(m+\bar{m})/\sqrt{2},\
\ \hat{\varphi}=i(m-\bar{m})/\sqrt{2}.
\end{equation}

\section{INVARIANTS}

In general, the quadratic Riemann invariant (Kretschmann scalar) is related
to the Weyl and Ricci invariants by%
\begin{equation*}
R_{\alpha \beta \mu \nu }R^{\alpha \beta \mu \nu }=C_{\alpha \beta \mu \nu
}C^{\alpha \beta \mu \nu }+2R_{\alpha \beta }R^{\alpha \beta }-\mathcal{R}%
^{2}/3.
\end{equation*}%
The quadratic invariants for $g^{\text{sSCH}}$ are, with $\Omega
=r_{0}^{2}/r^{3}$%
\begin{eqnarray}
R_{\alpha \beta \mu \nu }^{\text{sSCH}}R_{\text{sSCH}}^{\alpha \beta \mu \nu
} &=&48\frac{m_{0}^{2}}{r^{6}}+24\frac{m_{0}}{r^{3}}(3\Omega \sin \vartheta
)^{2}+R_{\alpha \beta \mu \nu }^{\text{sM}}R_{\text{sM}}^{\alpha \beta \mu
\nu },  \label{ssch-invars} \\
C_{\alpha \beta \mu \nu }^{\text{sSCH}}C_{\text{sSCH}}^{\alpha \beta \mu \nu
} &=&48\frac{m_{0}^{2}}{r^{6}}+24\frac{m_{0}}{r^{3}}(3\Omega \sin \vartheta
)^{2}+C_{\alpha \beta \mu \nu }^{\text{sM}}C_{\text{sM}}^{\alpha \beta \mu
\nu },  \notag \\
R_{\alpha \beta }^{\text{sSCH}}R_{\text{sSCH}}^{\alpha \beta } &=&\frac{243}{%
4}\Omega ^{4}\sin ^{4}\vartheta ,\ \ \ \mathcal{R}_{\text{sSCH}}^{2}=\frac{81%
}{4}\Omega ^{4}\sin ^{4}\vartheta ,  \notag
\end{eqnarray}%
and the quadratic invariants for $g^{\text{sM}}$ are 
\begin{eqnarray}
R_{\alpha \beta \mu \nu }^{\text{sM}}R_{\text{sM}}^{\alpha \beta \mu \nu }
&=&-3(\frac{6\Omega }{r})^{2}+2(\frac{6\Omega \sin \vartheta }{r})^{2}+\frac{%
9\times 99}{4}\Omega ^{4}\sin ^{4}\vartheta ,  \label{sm-invars} \\
C_{\alpha \beta \mu \nu }^{\text{sM}}C_{\text{sM}}^{\alpha \beta \mu \nu }
&=&-3(\frac{6\Omega }{r})^{2}+2(\frac{6\Omega \sin \vartheta }{r}%
)^{2}+108\,\Omega ^{4}\sin ^{4}\vartheta ,  \notag \\
R_{\alpha \beta }^{\text{sM}}R_{\text{sM}}^{\alpha \beta } &=&\frac{243}{4}%
\Omega ^{4}\sin ^{4}\vartheta ,\ \ \ \mathcal{R}_{\text{sM}}^{2}=\frac{81}{4}%
\Omega ^{4}\sin ^{4}\vartheta .  \notag
\end{eqnarray}%
For comparison, the Kerr invariant is%
\begin{eqnarray*}
R_{\alpha \beta \mu \nu }^{\text{kerr}}R_{\text{kerr}}^{\alpha \beta \mu \nu
} &=&C_{\alpha \beta \mu \nu }^{\text{kerr}}C_{\text{kerr}}^{\alpha \beta
\mu \nu } \\
&=&48\frac{m_{0}^{2}}{r^{6}}\frac{1-\alpha ^{2}\cos ^{2}\vartheta }{%
(1+\alpha ^{2}\cos ^{2}\vartheta )^{6}}(1-14\alpha ^{2}\cos ^{2}\vartheta
+\alpha ^{4}\cos ^{4}\vartheta )
\end{eqnarray*}%
where $\alpha =a/r$.

\section{POINCAR\'{E} INEQUALITY}

An inequality due to Poincar\'{e} \cite{Poi85} shows that $\Omega $ is
reasonable. Consider the equilibrium of a rotating, self-gravitating fluid
obeying Newtonian gravity. 
\begin{equation*}
\nabla ^{2}\phi =4\pi G\rho
\end{equation*}%
for uniform density $\rho $. The fluid is spinning with constant $\Omega $
about the $\hat{z}$ axis. The force on the liquid is 
\begin{equation*}
\vec{F}=-\nabla \lbrack \phi -(\Omega ^{2}/2)(x^{2}+y^{2})].
\end{equation*}%
For equilibrium, the total force on the boundary of fluid volume $V$ with
outward normal $\hat{n}$ must be 
\begin{eqnarray*}
\oint \vec{F}\cdot \hat{n}\,d^{2}x &\leq &0\text{\ \ }\Rightarrow \\
-\oint \nabla \lbrack \phi -(\Omega ^{2}/2)(x^{2}+y^{2})]\cdot \hat{n}%
\,d^{2}x &\leq &0\text{\ \ }\Rightarrow \\
-\int \nabla ^{2}[\phi -(\Omega ^{2}/2)(x^{2}+y^{2})]\,d^{3}x &\leq &0\text{%
\ \ }\Rightarrow \\
\int (4\pi G\rho -2\Omega ^{2})\,d^{3}x &\geq &0.
\end{eqnarray*}%
The mass, volume, and average density are 
\begin{equation*}
M=\int \rho d^{3}x\text{,\ \ }V=\int d^{3}x\text{, \ \ }\bar{\rho}=M/V
\end{equation*}%
and so 
\begin{equation}
\Omega ^{2}\leq 2\pi G\bar{\rho}.
\end{equation}%
It is simple to verify that $\Omega (r)=r_{0}^{2}/r^{3}$ satisfies Poincar%
\'{e}'s inequality for $r>r_{0}$ and quasilocal mass $M=r_{0}^{4}/r^{3}$ in
Eq.(\ref{m-komar}). The average density is $\bar{\rho}=M/(4\pi
r^{3}/3)=3r_{0}^{4}/4\pi $ and $\Omega ^{2}=r_{0}^{4}/r^{6}$.

\section{BONDI-SACHS METRIC}

Asymptotically flat systems are described by the Bondi-Sachs metric 
\begin{equation}
g_{\mu \nu }^{\text{BS}}dx^{\mu }dx^{\nu }=\frac{Ve^{2b}}{r}%
du^{2}+2e^{2b}dudr-r^{2}H_{AB}(dx^{A}-U^{A}du)(dx^{B}-U^{B}du),
\label{bsmetric}
\end{equation}%
where outgoing null hypersurfaces are labeled by $u$ and coordinates are $%
(x^{0}=u,\ x^{1}=r,\ x^{2}=\vartheta ,\ x^{3}=\varphi )$. The Bondi-Sachs
metric \cite{Sac62} extends Bondi's original metric \cite{Bon62} to include $%
\varphi $ dependence and has six independent functions $\{V,\ b,\
U^{\vartheta },\ U^{\varphi },\ y,\ q\}$, of ($u,r,\vartheta ,\varphi $).
The rays of each $u$ = const null surface are null geodesics $x^{\alpha }(r)$
with tangent $dx^{\alpha }/dr$ where $x^{1}=r$ is a luminosity parameter.
Coordinates $(\vartheta $, $\varphi )$ are constant along each ray. The
luminosity parameter is defined by $r^{4}\sin ^{2}\vartheta =\det (g_{AB})$ $%
=$ det$(r^{2}H_{AB})$ where 
\begin{equation*}
H_{AB}=\left[ 
\begin{array}{cc}
e^{2y}\cosh (2q) & \sinh (2q)\sin \vartheta \\ 
\sinh (2q)\sin \vartheta & \quad e^{-2y}\cosh (2q)\sin ^{2}\vartheta%
\end{array}%
\right] .
\end{equation*}%
The boundary conditions on the metric functions in the limit of future null
infinity $\mathcal{I}^{+}$ are 
\begin{equation*}
rU^{A}\rightarrow 0,\qquad b\rightarrow 0,\qquad y\rightarrow 0,\qquad
q\rightarrow 0,\qquad V/r\rightarrow 1.
\end{equation*}%
Notation was chosen to avoid confusion between Sachs metric functions and
Newman-Penrose spin coefficients: 
\begin{equation*}
2y=\gamma +\delta \ (\text{Sachs}),\quad 2q=\gamma -\delta \ (\text{Sachs}%
),\quad b=\beta \ (\text{Sachs}).
\end{equation*}%
The $u=$ const hypersurfaces have null geodesic tangent $l^{\alpha }\partial
_{\alpha }$ which is also hypersurface orthogonal as $l_{\alpha }dx^{\alpha
}=du$. The twist of $l^{\alpha }$ is zero and its expansion and shear are
given by 
\begin{equation}
\rho =-e^{-2b}/r,\ \ \sigma =-e^{-2b}[(\partial _{r}y)\cosh (2q)+i(\partial
_{r}q)]  \label{rhoand sig}
\end{equation}%
which follows from $r$ as a luminosity parameter, and where the phase of $%
\sigma $ is determined by the choice of tetrad orientation. The Bondi mass
aspect $M(u,\vartheta ,\varphi )$ is 
\begin{equation}
-2M=\Psi _{2}^{0}+\bar{\Psi}_{2}^{0}+\partial _{u}(\sigma ^{0}\bar{\sigma}%
^{0}),  \label{aspect}
\end{equation}%
where the zero superscript indicates the leading coefficient in a $1/r$
expansion. The Bondi mass is the two-surface integral of the mass aspect
over a topological two-sphere at $\mathcal{I}^{+}$%
\begin{equation}
M_{\text{Bondi}}=-\frac{1}{8\pi }\doint\limits_{S^{2}}[\Psi _{2}^{0}+\bar{%
\Psi}_{2}^{0}+\partial _{u}(\sigma ^{0}\bar{\sigma}^{0})]\,\sqrt{-g}%
d\vartheta d\varphi .  \label{bondimass}
\end{equation}

\end{document}